\documentstyle[12pt]{article}
\def\N{{\cal N}}

\def\O{{\cal O}}
\def\D{{\cal D}}

\def\cob{\delta}
\def\ep{\epsilon}

\def\Z{{\bf Z}}

\def\Tr{{\rm Tr}}

\def\hf{{1\over 2}}

\def\R{{\bf R}}
\def\o{\over}

\def\til#1{\widetilde{#1}}

\def\si{\sigma}
\def\Si{\Sigma}
\def\b#1{\overline{#1}}
\def\del{\partial}

\def\lap{\Delta}
\def\bra{\langle}
\def\ket{\rangle}
\def\lf{\left}
\def\ri{\right}
\def\riya{\rightarrow}

\def\la{\lambda}
\def\La{\Lambda}
\def\h#1{\widehat{#1}}
\def\bdel{\bar{\partial}}
\def\bt{\beta}
\def\ga{\gamma}
\def\Ga{\Gamma}
\def\al{\alpha}

\def\up{\uparrow}

\def\dag{\dagger}
\def\rt#1{\sqrt{#1}}

\def\CP{{\bf CP}}
\def\C{{\bf C}}

\def\Im{{\rm Im}}

\def\sitarel#1#2{\mathrel{\mathop{\kern0pt #1}\limits_{#2}}}

\def\bz{\b{z}}

\def\Det{{\rm Det}}
\def\up{\Upsilon}
\def\np#1#2#3{{ Nucl. Phys.} {\bf B#1} (#2) #3}

\def\pln#1#2#3{{Phys. Lett.} {\bf B#1} (#2) #3}

\def\pr#1#2#3{{ Phys. Rev.} {\bf D#1} (#2) #3}
\def\prl#1#2#3{{ Phys. Rev. Lett.} {\bf #1} (#2) #3}

\def\jhep#1#2#3{{JHEP} {\bf #1} (#2) #3}
\def\atmp#1#2#3{{Adv. Theor. Math. Phys.} {\bf #1} (#2) #3}
\def\hpt#1{{\tt hep-th/#1}}

\def\ijmp#1#2#3{{Int. J. Mod. Phys. } {\bf A#1} (#2) #3}

\def\lb       {\left( }
\def\rb       {\right) }

\def\lbb     {\left[ }
\def\rbb      {\right] }
\def\comma      { \, , }
\def\period     { \, . }
\def\semiket#1  { \, #1 \, \rangle \, }
\def\del        {  \partial  }
\def\half       {  {1\over 2}  }
\def\abs#1      {  \, \vert #1 \vert \,   }
\def\Im#1    { \, {\rm Im } \, #1  }
\def\Re#1    { \, {\rm Re}  \, #1  }

\def\bfR     { {\bf R}}
\def\bfZ     { {\bf Z}}
\def\bfC     { {\bf C}}
\def\delbar {\bar{\partial}}
\def\zbar   {\bar{z}}

\def\vecii#1#2      {  \left(\begin{array}{c}#1\\#2\end{array}\right)  }
\def\veciii#1#2#3   {  \left(\begin{array}{c}#1\\#2\\#3\end{array}\right)  }
\def\matrixii#1#2#3#4            {  \biggl( \begin{array}{cc}#1&#2\\#3&#4
                                       \end{array} \biggr) }
\def\matrixiii#1#2#3#4#5#6#7#8#9 {  \left(\begin{array}{ccc}#1&#2&#3\\
                                     #4&#5&#6\\#7&#8&#9\end{array}\right)  }
\def\eqabegin         {  \begin{eqnarray}  }
\def\eqaend           {  \end{eqnarray}  }
\def\nn               {  \nonumber  }

\def\sectionnumbering { \setcounter{equation}{0}
         \renewcommand{\theequation}{\arabic{section}.\arabic{equation}}}
\def\appendixnumbering { \setcounter{equation}{0}
         \renewcommand{\theequation}{\Alph{section}.\arabic{equation}}}
\def\mysection#1{ \addtocounter{section}{1} \setcounter{subsection}{0}
                 \sectionnumbering \par \bigskip
      \par \medskip
   \begin{center} {\sc \arabic{section} \quad  #1 } \end{center}
    \par }
\def\appsection#1{\addtocounter{section}{1} \setcounter{subsection}{0}
                 \appendixnumbering \par \bigskip \par \medskip
   \begin{center} {\sc \Alph{section} \quad  #1 } \end{center}
\par  }
\def\mysubsection#1{\addtocounter{subsection}{1}
      \par \bigskip \noindent  {\normalsize\it
      \arabic{section}.\arabic{subsection} \quad #1  }
   \par \medskip }

\def\csectionast#1    { \begin{center}
    {\large\bf #1  }   \end{center} \par \bigskip}
\renewcommand{\thefootnote}{\fnsymbol{footnote}}

\begin{document}
%
\def\papertitlepage{\baselineskip 3.5ex \thispagestyle{empty}}
\def\preprinumber#1#2#3{\hfill \begin{minipage}{4.2cm}  #1
              \par\noindent #2
              \par\noindent #3
             \end{minipage}}
\renewcommand{\thefootnote}{\fnsymbol{footnote}}
%
%
\papertitlepage
\setcounter{page}{0}
\preprinumber{KEK-TH-694}{UTHEP-424}{hep-th/0005152}
\baselineskip 0.8cm
\vspace*{2.5cm}
\begin{center}
{\large\bf Path Integral Approach to String Theory on $AdS_3$}
\end{center}
\vskip 8ex
\baselineskip 0.7cm
\begin{center}
        { Nobuyuki ~Ishibashi\footnote[2]
        {\tt ishibash@post.kek.jp}, 
       \  Kazumi ~Okuyama\footnote[3]{\tt kazumi@post.kek.jp} } \\
    {\it High Energy Accelerator Research Organization (KEK)} \\
    {\it Tsukuba, Ibaraki 305-0801, Japan} \\
         {\sl and} \\
        { Yuji  ~Satoh\footnote[4]{\tt ysatoh@het.ph.tsukuba.ac.jp}}  \\
    {\it Institute of Physics, University of Tsukuba} \\
    {\it Tsukuba, Ibaraki 305-8571, Japan}
\end{center}
\vskip 10ex
%
\baselineskip=3.5ex
\begin{center} {\bf Abstract} \end{center}
\par\smallskip\noindent
Using the path integral approach, we discuss the correlation functions of
the $SL(2,\bfC)/SU(2)$ WZW model, which corresponds to the string theory on
the Euclidean $AdS_3$. We obtain the two- and three-point functions for
generic primary fields in closed forms. By an appropriate change of
the normalization of the primary fields, our results coincide
with those by Teschner, which were obtained by using the bootstrap approach. 
The supergravity results are also obtained in the semi-classical limit.
\vskip 2ex
%
%
%
%
%
\vspace*{\fill}
\noindent
May 2000
\newpage
\renewcommand{\thefootnote}{\arabic{footnote}}
\setcounter{footnote}{0}
\setcounter{section}{0}
\baselineskip = 0.6cm
\pagestyle{plain}
\mysection{Introduction}
The three-dimensional anti-de Sitter space ($AdS_3$) is a simple
space-time with a constant negative curvature. From a group theoretical
point of view, it is nothing but the $SL(2,\bfR)$ group manifold.
Because of this simplicity, it provides useful testing grounds for
 investigating strings in curved space-time and
non-rational conformal field theories (see, for example, \cite{Petropoulos}
and references therein). Furthermore, $AdS_3$ is closely
related to various black hole geometries.
This implies that the string theory on $AdS_3$
offers a key to the quantum theory of black holes.

Besides these features, $AdS_3$ gives the simplest example of the
AdS/CFT correspondence \cite{Maldacena}-\cite{Witten}. This stimulated
the recent studies of the string theories on $AdS_3$ and its Euclidean
analog $SL(2,\bfC)/SU(2) = H_3^+$.
Since the S-dual configuration of the $D1/D5$-brane system does not have
the RR-field background, its near horizon geometry can be analyzed by
the standard world-sheet technique,
namely, the $SL(2,\bfR)$ or $H_3^+$ WZW model \cite{GKS}-\cite{SW}.
In particular, the authors of \cite{BORT,KS} showed that strings
in the bulk of $AdS_3$, or `short strings', can be treated
beyond the free field approach.

However, in spite of the importance of the $SL(2,\bfR)$ and $H_3^+$ WZW
models, it seems that these models are not yet completely understood.
For a precise understanding, one needs to clarify the fundamental properties
such as the true spectrum, modular invariance, fusion rules and unitarity.
For recent progress and discussions on these issues,
see \cite{Petropoulos},\cite{BDM}-\cite{HHS}.

In this paper, we concentrate on the $H_3^+$ WZW model corresponding
to the Euclidean $AdS_3$. This model has been studied by several approaches,
even before the AdS/CFT correspondence was proposed.
The early one used path integral \cite{Haba,Gawedzki},
and the correlation functions were obtained
for certain fields with non-negative half-integral $SL(2,\bfC)$ spins
\cite{Gawedzki}. The important fact is that the $H_3^+$ WZW model
allows us
the Lagrangian approach, which is impossible in the case of other WZW
models.
Later, for primary fields with generic spins, the correlation functions
and fusion rules were discussed based on the symmetry and bootstrap
\cite{Teschner1,Teschner2}. There are also arguments about
the correlation functions using the free field realization of the
$\widehat{sl_2}$ algebra \cite{CTT,GN} and the supergravity
approximation, e.g., \cite{GKP,Witten},\cite{MV}-\cite{KO}.
Taking these into account, the $H_3^+$ WZW model seems to be more tractable than
the $SL(2,\bfR)$ WZW model. Furthermore, since the precise formulation
of the AdS/CFT correspondence is given for the Euclidean $AdS$, $H_3^+$
seems to have a direct connection to this correspondence.

The aim of this paper is to discuss the correlation functions of the
primary fields with generic spins using the path integral approach.
In this approach, we can calculate the correlation functions
directly by a somewhat familiar method. Moreover, this enables us to 
discuss beyond the `free field approximation' \cite{GKS,GN,CTT}.
In the following, we first argue that,
by an appropriate definition of the correlation functions,
their calculation of is essentially reduced to that in \cite{Gawedzki}.
We then discuss in detail the cases of
 two- and three-point functions and obtain them in closed forms.
 The supergravity calculation is recovered in the semi-classical limit.
The results are also compared with those by Teschner
\cite{Teschner1,Teschner2}
and an exact agreement is found after a change of normalizations.
Thus, Teschner's approach and ours here are complementary to each other.
However, because of the advantages mentioned above, further
extensions of our approach may be possible. For example, we may be able to
calculate the correlation
functions of the energy-momentum tensor of the boundary CFT.

This paper is organized as follows. In section 2, we summarize
the $H_3^+$ WZW model. In section 3, we review the discussion of
\cite{Gawedzki} in some detail to make this paper self-contained.
In section 4, the general formalism is given for the calculation of
the correlation functions of generic primary fields. Section 5 is
devoted to the calculation of the two- and three-point functions.
In section 6, we compare our results with those from different approaches.
We conclude with a brief discussion in section 7. Some useful
integral formulas and properties of the $\up$-function, which appear
in section 6, are collected in appendix A and B, respectively.

\mysection{$H_3^+$ WZW model}
We begin  with a description of the conformal field theory
whose target space is the Euclidean $AdS_3$, namely,
$SL(2,\bfC)/SU(2) = H_3^+$ \cite{GKS}-\cite{KS},
\cite{Gawedzki}-\cite{Teschner2}.
After a brief summary, we introduce a spin $0$ primary field, which
appears  in a later discussion.
\mysubsection{Action and symmetry}
An element $g$ of $H_3^+$ is parametrized as
\eqabegin
   g  & = & e^{\ga\tau_+}e^{-\phi\tau_3}e^{\b{\ga}\tau_-}
      \ = \  \matrixii{e^{\phi}|\ga|^2+e^{-\phi}}{e^{\phi}\ga}
{e^{\phi}\b{\ga}}{e^{\phi}}
 \comma \label{gparam}
\eqaend
where $\gamma^* = \bar{\gamma}$; and
$\tau_\pm = (\tau_1 \pm i \tau_2)/2$ and  $\tau_3$ are Pauli
matrices.
The coset structure becomes manifest when $g$ is written as
\eqabegin
  g &=& h h^\dagger \comma \qquad h \ = \
   \matrixii{e^{-\phi/2}}{e^{\phi/2}\gamma}{0}{e^{\phi/2}} \ \in \
SL(2,\bfC)
  \comma
\eqaend
because $g$ is invariant under $h \to h u $ with $u \in SU(2)$.
In this parametrization, the isometry of $H_3^+$ is
\eqabegin
   g & \riya & g^A=AgA^\dag, \qquad A\in SL(2,\C)
  \period \label{global}
\eqaend

The conformal field theory with the target space $H_3^+$ is described by
the WZW action $S_{\rm WZW} (g(z))$. Substituting (\ref{gparam}) yields
\eqabegin
   S_{WZW} & = & {k\o\pi}\int d^2z\big(\del\phi\b{\del}\phi+
  e^{2\phi}\del\b{\ga}\b{\del}\ga\big)
  \period \label{SAdS}
\eqaend
Here, $k$ is the level of the WZW model and
$d^2z=d\si_1d\si_2$ with $z = \si_1 + i \si_2$.
$\del=\del_{z}$ and $\delbar=\del_{\bar{z}}$. The full theory is defined by
this action and the invariant measure,
\eqabegin
  \D g &=& \D\phi\D(e^{\phi}\ga)\D(e^{\phi}\b{\ga})
  \period \label{measure}
\eqaend

The above action and measure have left and right  affine symmetries
$ \widehat{SL}(2,\bfC)_L \times \widehat{SL}(2,\bfC)_R $, which act on $g(z)$
as $g(z) \to A(z) g(z) B^\dag(z)$ with  $A(z),B(z) \in SL(2,\bfC)$.
In order to keep $g$ an element of $H_3^+$, they need to be related 
to each other by $A(z) = B(z)$. Thus, the symmetry of the model is
$ \widehat{SL}(2,\bfC) \times \widehat{\overline{SL}}(2,\bfC)$.
In particular, the global
symmetry corresponds to a constant matrix $A$ and given by
(\ref{global}). The currents of this global symmetry are
represented by
\eqabegin
  && J^-_0 = \del_\gamma \comma \quad
   J^3_0 = \gamma \del_\gamma - \half \del_\phi \comma \quad
    J^+_0 = \gamma^2 \del_\gamma - \gamma\del_\phi - e^{-2\phi}
 \del_{\bar{\gamma}}
  \comma
\eqaend
and similar expressions with bars.

The action (\ref{SAdS}) can be rewritten by introducing the auxiliary
fields $\beta$ and $\bar{\beta}$ \cite{GKS}. The resultant action
looks like an action for the free fields $\phi$, $\beta$-$\gamma$ and
$\bar{\beta}$-$\bar{\gamma}$, except for the term
$\beta \bar{\beta} e^{-2\phi}$. This additional term drops out in the region
$\phi \to \infty$, which corresponds to the boundary of $H_3^+$.
Thus, near to the boundary of $H_3^+$ the free field approach is applicable,
but it is not completely clear to what extent one can use this approach
in a generic region. Regarding this issue, see \cite{BORT,KS,GN}.
In our approach based on the full Lagrangian, we do not have such
subtleties.
\mysubsection{Primary fields}
The primary fields of the model form the representations of the global
$ SL(2,\bfC)$. These representations are well
organized by
introducing auxiliary coordinates $(x, \bar{x})$ \cite{GGV}.
They are interpreted
as the coordinates of the boundary CFT in the AdS/CFT correspondence
\cite{BORT}.
Using these, the spin $j$ primary field is given by
\eqabegin
  \Phi_j \lb g(z),x \rb & = &
   \lf[(1,-x)g\lf(\matrix{1\cr -\b{x}}\ri)\ri]^{2j} \nn \\
   &=&
   \lb |\ga(z)-x|^2e^{\phi(z)}+e^{-\phi(z)} \rb^{2j}
  \period \label{Phijcl}
\eqaend
Note that one cannot separate the left and right sectors in this expression.
This is because the left and right symmetries are related to each other.
By expanding $\Phi_j$ in terms of $(x,\bar{x})$ as
\eqabegin
  \Phi_j &=& \sum_{m,\bar{m}} x^{j-m} \bar{x}^{j-\bar{m}}
  \Phi^{j}_{m \bar{m}}
  \comma
\eqaend
one obtains primary fields with definite eigenvalues $(m,\bar{m})$ of
$(J^3_0,\bar{J}^3_0)$.
The range of $m$ and $ \bar{m}$ depends on
the value of $j$. For example, $\Phi_{1/2}$ is expanded as
\eqabegin
    \Phi_{1/2} &=& ( |\gamma|^2 e^\phi + e^{-\phi} ) -x(\bar{\gamma}e^\phi)
       - \bar{x}(\gamma e^\phi) + x\bar{x} e^\phi
     \period \label{Phi1/2}
\eqaend

It is straightforward to check that the action of the $SL(2,\bfC)$ currents
on  $\Phi_j$ gives
\eqabegin
 J^a_0\Phi_j(g,x) & = & -D^a\Phi(g,x)
  \comma \label{primary}
\eqaend
where
\eqabegin
  && D^- = \del_x \comma \quad D^3 = x\del_x-j \comma \quad
      D^+ = x^2\del_x-2jx
  \period \label{Da}
\eqaend
In other words, $\Phi_j$ transforms under the global transformation
(\ref{global}) as
\eqabegin
  (R_A\Phi_j)(g,x) & = & \Phi_j(g^{A^{-1}},x) \ = \ |cx+d|^{4j}\Phi_j(g,Ax)
   \label{RAPhi}
\eqaend
with
\eqabegin
  Ax & = & {ax+b\o cx+d} \comma \qquad
  A \ = \ \matrixii{a}{b}{c}{d} \ \in \ SL(2,\bfC)
  \period
\eqaend

In the discussions of the $H_3^+$ WZW model,
various $SL(2,\bfC)$ representations
appear. An important class is called the principal continuous series.
This class of representations is unitary, and has spin
$ j = -1/2 + i \rho $ $(\rho \in \bfR)$.\footnote{
$J_0^3$ and $\bar{J}_0^3$ take
$m= (ip+n)/2, \bar{m} = (ip-n)/2$ with $p \in \bfR \comma n \in \bfZ$.
These are different from the corresponding representation of $SL(2,\bfR)$
with the same $j$ because $m$ and $\bar{m}$ are real in this case.
}
The space of the square-integrable functions on $ H_3^+ $
is decomposed into these representations ${\cal H}_{-1/2 + i \rho}$
\cite{GGV}:
\eqabegin
   L^2(H_3^+,dg) & \cong & \int_{\rho > 0}^\otimes d \rho \, \rho^2\,
   {\cal H}_{-1/2 + i \rho} \period \label{L2}
\eqaend
A class of the representations with $ j \leq -1/2 $
appears in the discussion of the AdS/CFT correspondence \cite{GKS}-\cite{KS}.
This is an analog of the discrete series representations of $SL(2,\R)$.
When the spin is a non-negative half-integer, $\Phi_j$ is expanded into
a finite sum of $\Phi^j_{m,\bar{m}}$ as in (\ref{Phi1/2}).
This case appears in relation to
the $SU(2)$ WZW model \cite{Gawedzki}.

Since spin $j$ is just the label of the second Casimir of the
$SL(2,\bfC)$, i.e.,
$-j(j+1)$, the representation with $j$ and that with $-j-1$
are equivalent.
This appears to be obvious for the principal continuous series because of
the relation $-j-1 = j^* $.
In general, the primary fields $\Phi_j$ and $\Phi_{-j-1}$
are classically related by
\eqabegin
   \Phi_j(g,x) & = & {2j+1\o\pi}\int d^2y \ |x-y|^{4j}  \Phi_{-j-1}(g,y)
   \period \label{j-j-1}
\eqaend
However, this expression for a generic $j$ may be modified at the values
$j \in \bfZ/2$. This phenomena is called `resonance' in \cite{KS}.
\newpage
\mysubsection{Spin $0$ primary}
It turns out that the spin $0$ primary field appears in the definition
of our correlation functions. An obvious `spin $0$  primary field' is
just a constant $\Phi_0$. In addition to this,
there exists a non-trivial spin $0$ field.
In fact, because of  the formula
\eqabegin
  f(\ga)\del_{\ga}^n\cob(\ga-x)=\sum_{l=0}^n(-1)^l\lf(\matrix{n\cr l}\ri)
\del_x^lf(x)\del_{\ga}^{n-l}
\cob(\ga-x)
 \comma
\eqaend
the operator
\eqabegin
  \h{\Phi}_0(g,x) & = & \sum_{n=0}^{\infty}{(-1)^n\o
 n!(n+1)!}e^{-2(n+1)\phi}\del_{\ga}^n\del_{\bar{\ga}}^n\cob^2(\ga-x)
\eqaend
satisfies (\ref{primary}) with $j = 0$. Up to a coefficient and a constant
term, this is also obtained by
taking the limit $j \to 0$ in (\ref{j-j-1}), namely,
\eqabegin
  \h{\Phi}_0 & \sim & \int d^2x \, \Phi_{-1}(x)
  \period
\eqaend
In deriving this, we need the integral formula
(\ref{singleint}) in appendix A and
\eqabegin
  \lim_{\ep\riya 0}{|x|^{-2+2\ep}\o\Ga(\ep)} & = & \pi\cob^2(x)
   \comma \label{GS}
\eqaend
in \cite{GS}.
This implies that the right-hand side of (\ref{j-j-1}) with
$j \to 0$ is different from $\Phi_0$ in (\ref{Phijcl}).
This is an example of the `resonance' mentioned above.

Since $\Phi_{-1}$ corresponds to the dimension $(1,1)$ operator of the
boundary CFT,
$\h{\Phi}_0$ is regarded as its vertex operator.
Moreover, at least semi-classically $\Phi_{-1}$ satisfies  \cite{KS}
\eqabegin
  \del_{\bar{z}}\Phi_{-1} & = & \del_{\bar{x}}\big(\bar{J}\Phi_{-1}\big)
   \comma \nn \\
  \bar{J}\Phi_{-1}& = & \del_{\bar{z}}\La
  \comma  \\
   \Phi_{-1} \Phi_j & \sim & (\mbox{regular terms in } z, \zbar)
   \comma \nn
\eqaend
and similar equations with $\del_z, \del_x$ and $J$.
Here $\bar{J}(x,z)= 2x \bar{J}^3(z) -\bar{J}^+(z) -x^2 \bar{J}^-(z)$,
$\bar{J}^a(z)$ are the $\widehat{sl_2}$ currents, and
$\Lambda(g,x)$ is a certain function. $J$ is defined similarly.
The first two equations  further imply that
\eqabegin
 \del_{\bar{z}}\h{\Phi}_0 & = &\del_{\bar{z}}\int d^2x\del_{\bar{x}}\La
  \ \sim \ 0
  \period
\eqaend
Thus, we see that $\h{\Phi}_0$ behaves as the identity (constant) operator
on the world-sheet.  Although this argument is based on the semi-classical
analysis in \cite{KS}, we will see that $\h{\Phi}_0$ actually behaves as
the identity. This supports the argument in \cite{KS} conversely
from the point of view of the full quantum theory.
Note that the identity operator of the boundary theory is
\eqabegin
  I & = & \int d^2zJ\bar{J}\Phi_{-1}
  \period
\eqaend
\mysection{Review of path integral approach}
The $H_3^+$ WZW model was discussed early in \cite{Haba}, and
it was found that the functional integral of certain correlation
functions can be performed. Later, such an argument was further
developed by Gaw{\c e}dzki in relation to $G/H$ coset models
\cite{Gawedzki}.

In the next section, we discuss the
correlation functions of the primary fields $\Phi_j$ with generic $j$.
We argue that the calculation of these correlation functions 
can be reduced to
that of certain correlation functions with $j \in \bfZ_{\geq0}/2$,
which has been discussed by Gaw{\c e}dzki \cite{Gawedzki}.
Thus, we first review his discussion.
It is understood that all of the spins are non-negative half-integers
in this section.

In the study of the $H_3^+$ WZW model, a difficulty  often arises
from the non-compactness of $H_3^+$. In the path integral
approach, this typically appears as the problem of zero-modes
and requires a careful treatment of them in the definition of
the correlation functions.

A naive definition of a correlation function may be
\eqabegin
   \Big\bra \O \Big\ket &\sim& \int \D g \,e^{-S_{WZW}[g]} \O
   \period \label{naive}
\eqaend
In a compact case such as $U(1)$, the zero-mode part in the functional
integral $\int \D g$
picks up the invariant part of $\O$. However, in the non-compact case
the zero-mode integral diverges generically. The  prescription
in \cite{Gawedzki} is: (i) choose $\O$ which is already invariant under
the global symmetry, and (ii) fix the zero-mode integral by inserting
a delta function $\delta (g(z_0)-g_0)$ in the functional integral, 
where $g_0$ is a constant element of $SL(2,\bfC)$.
The delta function here maintains both the independence of $z_0$ and
 invariance under $\widehat{SL}(2,\bfC)$: since $\O$ is invariant
under $SL(2,\bfC)$, the insertion of
$V^{-1}= V^{-1} \int dA \delta(g^A(z_0)-g_0)$ with
$A \in SL(2,\bfC)$  and $V$ the volume of $SL(2,\bfC)$ gives
the delta function after the integration over  $A$.
Note that the integration over $A$ splits into those over $H_3^+$
and $SU(2)$ since $g^A = (Ah)(Ah)^\dag$.
Hence, instead of (\ref{naive}), the correlation function is defined by
\eqabegin
     \Big\bra \O \Big\ket &=& \frac{1}{Z_0}
      \int \D g \,\delta(g(z_0) -g_0) \ e^{-S_{WZW}[g]} \O
   \comma \label{true}
\eqaend
with $Z_0 = \int \D g \,\delta(g(z_0) -g_0) \ e^{-S_{WZW}}$ and
$\O$ invariant under the global $SL(2,\bfC)$.
In the parametrization (\ref{gparam}), the delta function takes the form
\eqabegin
   \delta(g(z_0)-g_0) &=& \delta^2
            \Big(e^{\phi(z_0)}(\gamma(z_0)-\gamma_0) \Big)
                          \delta(\phi(z_0) -\phi_0) \nn \\
       & = & e^{-2\phi(z_0)}
               \delta^2(\gamma(z_0)-\gamma_0 ) \delta(\phi(z_0) -\phi_0)
   \period \label{delta}
\eqaend

The simplest example of the correlation functions is the two-point function
of the spin $j = 1/2$ field. In this case, the invariant operator is
given by\footnote{
From the geometrical point of view, this represents the distance in
$H_3^+$:
$
  \half \Tr (g_1 g_2^{-1}) = u_{12}+ 1 \ = \ \cosh \sigma_{12}
  \comma
$
where $u_{12}$ and $\sigma_{12}$ are the chordal and geodesic
distances of $H_3^+$, respectively.
}
\eqabegin
  \O_2^G(j=1/2) &=& \Tr \lb g(z_1) g^{-1}(z_2) \rb  \\
       &=& e^{\phi(z_1)-\phi(z_2)} + e^{\phi(z_2)-\phi(z_1)}
           + e^{\phi(z_1)+\phi(z_2)} |\gamma(z_1)-\gamma(z_2)|^2
   \period  \nn
\eqaend

Since the action is bi-linear in $\gamma$, the functional integral
over $\gamma$ is Gaussian and can be carried out. The propagator
is then\footnote{Here, we use $\gamma$ for $\gamma -\gamma_0$ for simplicity.
We use a similar kind of abuse of notation for $\phi$ in the following.
Since we will calculate expectation values of quantities invariant under
the global $SL(2,\C)$, we do not have to be careful about such notations.}
\eqabegin
  \Big\bra \b{\ga}(z)\ga(w) \Big\ket & = & \int {d^2y\o\pi k}
e^{-2\phi(y)}\lf({1\o \b{y}-\b{z}}+\hf\bdel\si(y)\ri)
  \lf({1\o y-w}+\hf\del\si(y)\ri)
 \comma \label{Ggamma}
\eqaend
where $\sigma $ is the conformal factor of the metric
$ g_{ab} = e^{\sigma} \delta_{ab}$ in the conformal gauge.  This satisfies
\eqabegin
-{k\o\pi}\del_{\b{z}}e^{2\phi(z)}\del_{z}\Big\bra \b{\ga}(z)\ga(w)
 \Big\ket
&=& \cob^2(z-w)- \nu(z)
 \comma
\eqaend
with
\eqabegin
  \nu(z) & = & \frac{1}{8\pi}\rt{g}R \ = \ - \frac{1}{2\pi}\del\bdel\si
  \comma
\eqaend
and $\int \nu(z) = 1$. $R$ is the world-sheet curvature.
The $\sigma$-dependence in (\ref{Ggamma})
disappears in the actual calculation, since it turns out that
one needs only the $\sigma$-independent combination
\eqabegin
  \lf\bra \Big(\b{\ga}(z_1)-\b{\ga}(z_2)\Big)
      \Big(\ga(z_3)-\ga(z_4)\Big) \ri\rangle
 = \int {d^2y\o \pi k}
e^{-2\phi(y)}{(\b{z}_1-\b{z}_2)(z_3-z_4)\o
       (\b{y}-\b{z}_1)(\b{y}-\b{z}_2)(y-z_3)(y-z_4)} \period
\eqaend
When the world-sheet curvature is concentrated on $ z= \h{z}_0 $,
$\si(z)$ is given by $-4\log|z-\h{z}_0|$.  In other words,
\eqabegin
ds^2=e^{\si(z)}|dz|^2={|dz|^2\o |z-\h{z}_0|^4}=\lf|d\lf({1\o
z-\h{z}_0}\ri)\ri|^2
\period
\eqaend
Then, the propagator further satisfies
$\Big\bra \b{\ga}(\h{z}_0)\ga(w) \Big\rangle = 0$. This is consistent with
the boundary condition following from the delta function (\ref{delta})
if $z_0 = \h{z}_0$. Thus, this choice of the propagator implies that
$z_0$ is the point of the support of the curvature.
However, this fact is not important in the actual calculation,
since the choice of $\h{z}_0$ is arbitrary, and hence so is $z_0$.
In fact, the expressions of the correlators
(without the $\sigma$-dependent part) turn out to be independent of
$z_0$ and hence one need not necessarily take $z_0 \to \h{z}_0 $.
This is understood as the remnant of the original $SL(2,\bfC)$ invariance.

Taking the measure (\ref{measure}) into account,
one finds that the $\gamma$-integration gives the Jacobian,
\eqabegin
   & & \Det^{'-1}(e^{-\phi-\hf\si}\b{\del}e^{2\phi}\del e^{-\phi-\hf\si})
   \period
\eqaend
By the standard procedure, this Jacobian is found to be
\cite{Gerasimov,Kallosh}
\eqabegin
 & &
   \exp\lf({1\o\pi}\int d^2z\, 2\del\phi\b{\del}\phi+{\phi\o4}\sqrt{g} R
+{1\o12}\del\si\bdel\si\ri)
   \comma
\eqaend
where we have dropped
$\det^{\prime -1}\del\delbar$, which is canceled by $Z_0$.

Consequently, the resultant effective action for $\phi$ becomes
\eqabegin
  S_{\phi} & = & {1\o\pi}\int d^2z\lf[(k-2)\del\phi\b{\del}\phi
                  -{\phi\o4} \sqrt{g} R \ri]
  \period \label{Sphi}
\eqaend
Using this, the two-point function is written as
\eqabegin
  \Big\langle \O_2^G(j=1/2) \Big\rangle &\sim & \int \D \phi
  \delta (\phi(z_0) -\phi_0) e^{-2\phi(z_0)} e^{-S_\phi} \label{O2mid}  \\
   && \quad \times \lb e^{\phi(z_1)-\phi(z_2)} + e^{\phi(z_2)-\phi(z_1)}
           + e^{\phi(z_1)+\phi(z_2)}
        \Big\langle |\gamma(z_1)-\gamma(z_2)|^2 \Big\rangle \rb
    \period \nn
\eqaend
The $\phi$-charge in this expression is neutral because the contribution
form the anomaly term $\sqrt{g}R$ is canceled with that from
$e^{-2\phi(z_0)}$.

The last ingredient to complete the calculation is the propagator of $\phi$.
The choice in \cite{Gawedzki} is
\eqabegin
    && \Big\bra \phi(z) \phi(w) \Big\rangle
      =  -b^2\log|z-w| + G_{\sigma}  \comma \nn \\
   && G_{\sigma} = -\frac{1}{4} b^2 \lb \sigma(z) + \sigma(w)
    + \frac{1}{2\pi} \int d^2z \sigma \del \delbar \sigma \rb
    \comma \label{Gphi}
\eqaend
with
\eqabegin
b^2\equiv {1\o k-2} \period
\eqaend
This propagator satisfies
\eqabegin
  -\frac{2}{b^2\pi} \del_{\zbar} \del_z \Big\bra \phi(z) \phi(w) \Big\rangle
   &  = & \delta^2(z-w) - \nu(z)
   \comma
\eqaend
and $\Big\bra \phi(\h{z}_0) \phi(w) \Big\rangle  = 0$.
The latter is again consistent with the boundary
condition if $z_0 = \h{z}_0$.
Note that, with this choice,
the curvature term in (\ref{Sphi}) does not contribute to the following
calculation, since $\int \big\bra \phi \phi \big\rangle
\del \delbar \sigma = 0$.

An important point here is that the $\phi$-integration gives divergent
factors through the self contraction of $\phi$. We regularize this
divergence by the point splitting method. Namely, we replace
$\Big\bra \phi (z)\phi (z)\Big\rangle$ by
\eqabegin
  \Big\bra \phi(z) \phi(z+\Delta z) \Big\rangle
  = -b^2 \log \epsilon
  - \frac{b^2}{8\pi} \int \sigma \del \delbar \sigma \comma
\eqaend
where  $\epsilon$ is the infinitesimal UV cut-off and
\eqabegin
    \epsilon & = & {\rm dist}(z,z+\Delta z)
    \ = \ e^{\hf\sigma(z)}|\Delta z|
  \period
\eqaend
The strongest divergence comes from the term including
$\Big\langle |\gamma(z_1)-\gamma(z_2)|^2 \Big\rangle$. In fact, it diverges
as $\epsilon^{-5b^2}$, since $e^{a\phi}$ in the correlator gives
$\epsilon^{-a^2b^2/2}$. This requires
the multiplicative renormalization $Z_0 \to \epsilon^{-2b^2}Z_0$,
which cancels the divergence from $e^{-2\phi(z_0)}$, and
\eqabegin
  \O_2^G(j=1/2) & \to & \epsilon^{-4\Delta_{1/2}} \O_{2}^G(j=1/2)
  \period
\eqaend
Here, $\Delta_j$ is the expected scaling dimension of the spin $j$ field,
\eqabegin
  \Delta_j &=& -b^2 j(j+1)
  \period
\eqaend
Because of this renormalization, the first and second terms in (\ref{O2mid})
disappear. This is the simplest example of the general rule:
for non-negative half-integral $j$, only the term with the highest power of
$\gamma$ survives the renormalization. This is confirmed by simple counting.

In all, omitting the factor including $\sigma$'s which  have
the support only at $\h{z}_0$, one arrives at
\eqabegin
  && \Big\langle \O_2^G(j=1/2) \Big\rangle \\
   && \qquad  =  |(z_0-z_1)(z_0-z_2)|^{2b^2} |z_1-z_2|^{2-b^2}
    \int d^2y \ |y-z_0|^{-4b^2} |(y-z_1)(y-z_2)|^{-2+2b^2}
  \period \nn
\eqaend
By simple changes of variables, one confirms that
this is independent of $z_0$, as expected.
The $\si$-dependence is
found to be ${\cal A}_2(\si,z_1,z_2,\hf,\hf)$,
which is defined by \cite{Gawedzki}
\eqabegin
{\cal A}_n(\si,z_a,j_a)
\ = \ \exp\lf({c\o24\pi}\int d^2z\del\si\bdel\si
-\sum_{a=1}^n\lap_{j_a}\si(z_a)\ri)
\eqaend
with
\eqabegin
c=2+1+6b^2={3k\o k-2} \period
\eqaend
This $\si$-dependence is canceled by the
internal CFT and $b$-$c$ ghosts when we consider the critical string theory.

An interesting consequence in this calculation is that
the Coulomb-gas picture of the free field approach naturally appeared:
the anomaly term $\sqrt{g}R$ corresponds to the charge at infinity
(when $\h{z}_0 \riya \infty$) and the $\gamma$-propagator looks like
the screening operator. Thus, the calculation here seems similar to
that of the free field approach. However, the precise relationship does not
seem to be completely clear.

The generalization of the above discussion to a generic
$j \in \bfZ_{\geq 0}/2$ is straightforward.
In such a case, the invariant combination of the two-point function
is
\eqabegin
  \O_2^G(j) &=& P^{2j} \lbb \Tr (g(z_1)g^{-1}(z_2))\rbb
  \comma
\eqaend
where  $P^{2j}(x)$ is a polynomial of order $2j$ with coefficient
$1$ at $x^{2j}$. Repeating a similar procedure, one finds
the renormalization
\eqabegin
   \O_2^G(j) &\to& \epsilon^{-4\Delta_j} \O_2^G(j)
   \comma \label{O2ren}
\eqaend
and the term which survives the renormalization,
\eqabegin
  \tilde{\O}_2^G(j) &=& e^{2j \bigl( \phi(z_1)+\phi(z_2)\bigr) }
   |\gamma(z_1)-\gamma(z_2)|^{4j}
  \period \label{tilO2G}
\eqaend

Finally, we consider the three-point function for spins $j_1,j_2,j_3$
with
\eqabegin
  && j_1+j_2+j_2 \, \in \, \bfZ  \comma \quad
    |j_1-j_2| \, \leq \, j_3 \, \leq \, j_1 + j_2
   \period \label{j1-j3}
\eqaend
These conditions assure that $j_{ab}$ defined by
$j_{12} = j_1+j_2-j_3$ and similar expressions are also non-negative
integers. In this case,
the invariant combination is obtained by using an invariant tensor,
the explicit form of which is found in \cite{Gawedzki}.
Then, similarly to the above, one finds the renormalization factor
$\epsilon^{-2(\Delta_{j_1}+\Delta_{j_2}+\Delta_{j_3})}$ and
the relevant term after the renormalization, 
\eqabegin
  \tilde{\O}_3^G(j_a) &=& e^{2j_1\phi(z_1)+2j_2\phi(z_2)+2j_3\phi(z_3)}
    \label{tilO3G} \\
          && \quad \times  |\gamma(z_1)-\gamma(z_2)|^{2j_{12}}
            |\gamma(z_2)-\gamma(z_3)|^{2j_{23}}
            |\gamma(z_3)-\gamma(z_1)|^{2j_{31}}
  \period \nn
\eqaend
The invariants (\ref{tilO2G}) and (\ref{tilO3G}) will appear
in a later discussion.
\mysection{Correlation functions of primary fields}
In the previous section, we reviewed the calculation in \cite{Gawedzki}
of certain correlation functions for non-negative half-integral spins.
We now move on to the discussion of the correlation functions of the
primary field $\Phi_j(g,x)$ with generic $j$.
\mysubsection{Definition of correlation functions}
In section 3, we saw that a careful treatment of zero-modes
is necessary because of the non-compactness of $H_3^+$.
Taking this into account, we define the correlation function of $\Phi_j$ by
\eqabegin
  \lf\bra \prod_{a=1}^{n}\Phi_{j_a}(g(z_a),x_a) \ri\ket
  & = & \frac{1}{Z} \int \D g e^{-S_{WZW}[g]} \h{\Phi}_0
\O_n(g(z_a),j_a,x_a)
   \comma \label{Phicrfn}
\eqaend
with $Z=\int \D g e^{-S_{WZW}} \h{\Phi}_0$.
Here $\O_n$ is the invariant part of
$\prod_{a=1}^{n}\Phi_{j_a}(g(z_a),x_a)$,
which will be determined in the next subsection.
Since $\O_n$ is invariant under the global symmetry and $\del_\gamma =
J^-_0$,
we find that, in the correlator, $\h{\Phi}_0$ is reduced to
\eqabegin
  \h{\Phi}_0 &\to & e^{-2\phi(z_0)} \delta^2 (\gamma(z_0)-x_0)
  \comma
\eqaend
and hence $\D g \h{\Phi}_0$ to
\eqabegin
  \D g \h{\Phi}_0 &\to & \D (e^\phi \gamma)\D (e^{\phi}\b{\ga})
             d\phi_0\D'\phi \delta^2(\gamma(z_0) - x_0) e^{-2\phi(z_0)}
   \period
\eqaend
Here, we have separated the measure of $\phi$  to its zero-mode part $d\phi_0$
and non-zero-mode part $\D' \phi$. Since the $\O_n$ is invariant
under the global $SL(2,\C)$,  the integrand in (\ref{Phicrfn})
does not depend on the zero-mode of $\phi$.
The divergent volume coming from the integration of
$\phi_0$ is canceled by the same factor in $Z$.
The $SL(2,\bfC)$ invariance and the independence of $z_0$ and $x_0$ follow
from the fact that $\h{\Phi}_0$
behaves as the identity operator on the world-sheet.

Thus, the treatment of zero-modes here seems to be 
almost the same as that discussed in the previous section.
In fact, we find that it is equivalent.
In (\ref{delta}), an additional
delta function $\delta(\phi(z_0)-\phi_0)$ was inserted instead of
performing the zero-mode integral and dividing by its volume.
This delta function imposed a boundary
condition on $\phi(z)$. In turn, this boundary condition
imposed the choice of the propagator (\ref{Gphi}) and the condition
$z_0 \to \h{z}_0$.
However, such a difference does not matter. This is because
(i) in the following calculation we use the same Green functions
(\ref{Ggamma}) and (\ref{Gphi})
(this is indeed a consistent choice), and (ii) though logically
one should take $z_0 \to \h{z}_0$ in the last step in the previous section,
$z_0$ disappears in the actual calculation
as discussed in the previous section.

Although our treatment of zero-modes and Gaw{\c e}dzki's are essentially
equivalent, the use of $\h{\Phi}_0$ has an advantage. Since $\h{\Phi}_0$
transforms as a primary field under $SL(2,\C)$, it is a conformal
field. Thus, the conformal property of the correlation function is manifest.
However $\delta(\phi(z_0)-\phi_0)$ does not transform properly under
$SL(2,\C)$,
so its inclusion in the path integral makes the transformation property
of the correlation functions apparently unclear.

The zero-mode integral is convergent for some correlation functions
such as those of the primary fields in the principal continuous series.
In such cases the prescription in eq.(\ref{Phicrfn}) coincides with the 
usual definition, because the zero-mode integral,
 which was fixed by the insertion of $\h{\Phi}_0$, is essentially
recovered by the projection defined in the next subsection.
\mysubsection{$SL(2,\bfC)$ projection}
To proceed further, we need to determine the invariant part
of $\prod_{a=1}^{n}\Phi_{j_a}(g(z_a),x_a)$. This is achieved by the
following projection:\footnote{
 Precisely, the integral is over $PSL(2,\bfC)$ in the later calculations.
}
\eqabegin
  \O_n(g(z_a),j_a,x_a) &= &
  \N\int_{SL(2,\C)}dA\prod_{a=1}^n(R_A\Phi_{j_a})(g(z_a),x_a)
  \comma \label{SL2projection}
\eqaend
where $\N$ is a normalization factor.
Indeed, if the integral over $SL(2,\C)$ is convergent,
$\O_n$ is invariant under $SL(2,\C)$:
\eqabegin
  \O_n(g^A(z_a)) & = &\O_n(g(z_a))
   \period
\eqaend

This projection can be performed explicitly. As the simplest example,
let us first consider the $n=2$ case. From (\ref{RAPhi}), it follows that
\eqabegin
  && \O_2(g(z_a)) =
\N\int{d^2ad^2cd^2d\o|c|^2}|cx_1+d|^{4j_1}|cx_2+d|^{4j_2}
\Phi_{j_1}(g(z_1),Ax_1)\Phi_{j_2}(g(z_2),Ax_2)
 \period \quad \
\label{acd}
\eqaend
The change of variables $(a,c,d)\riya (y,\la, w)$ with
\eqabegin
  && y=Ax_1 \comma \quad \la=cx_1+d \comma  \quad w = \frac{1}{\lambda c}
   \comma
\eqaend
gives
\eqabegin
   \O_2(g(z_a))&= & \N\int d^2yd^2\la d^2c|\la|^{4j_1}|cx_{12}-\la|^{4j_2}
      \Phi_{j_1}(g(z_a),y)\Phi_{j_2}(g(z_2),y+{x_{12}\o\la(cx_{12}-\la)})
   \nn \\
     &\simeq & \N|x_{12}|^{4j_2}\int d^2yd^2\la d^2c|\la|^{4j_1}|c|^{4j_2}
     \Phi_{j_1}(g(z_1),y)\Phi_{j_2}(g(z_2),y+{1\o\la c}) \label{adylam} \\
    &= &\N|x_{12}|^{4j_2}\int d^2yd^2\la
d^2w|\la|^{-2+4(j_1-j_2)}|w|^{-4-4j_2}
        \Phi_{j_1}(g(z_1),y)\Phi_{j_2}(g(z_2),y+w) \nn
    \comma
\eqaend
were $x_{12} = x_1 -x_2$.
We will use similar notations in the following.
In going from the first line to the second line above,
we have been sloppy about the treatment of the singular
parameter region $x_{12}\to 0$. As we will show, a careful treatment of such
a region gives an additional contribution to $\O_2$.
Let us define $\O_2^\prime$ to be the quantity in the last line of
eq.(\ref{adylam}).
By further renaming the variables, $\la=y_3$, $y=y_1$ and $y+w=y_2$,
$\O_2^\prime (g(z_a))$ becomes
\eqabegin
  \O'_2(g(z_a))&=& \N|x_{12}|^{4j_2}
   \int\prod_{a=1}^3d^2y_a|y_3|^{-2+4(j_1-j_2)}|y_{12}|^{-4-4j_1}
   \Phi_{j_1}(g(z_1),y_1)\Phi_{j_2}(g(z_2),y_2) \nn  \\
   &= &\pi^2\N i\cob(j_1-j_2)|x_{12}|^{4j_2}\int d^2y_1d^2y_2
   |y_{12}|^{-4-4j_1}\Phi_{j_1}(g(z_1),y_1)\Phi_{j_1}(g(z_2),y_2)
  \period \nn \\
   && \label{O2g}
\eqaend

To obtain the second line, the spins have to take the values of the
principal
continuous series $j_a = -1/2 + i \rho_a$ and
$i\delta (j_1-j_2)$ should be understood as  $\cob(\rho_1-\rho_2)$.
For other values, the integral
over $y_3$ is not well-defined. Thus, in such cases we `continue'
the expression of the second line to generic $j$. We will find that this
prescription is consistent with the three-point function.
In other words, we obtain the same result by (i) calculating the two-point
functions for $j_a = -1/2 + i \rho_a$ from the three-point functions
and (ii) continuing the final expression
to generic spins. In any case,  it is straightforward
to check that (\ref{O2g}) is invariant under the $SL(2,\bfC)$
transformation.
Note that the above  integral is nothing but
$\int d^2x \Phi_{j_1}(g(z_1),x) \Phi_{-j_1-1}(g(z_2),x)$.

In (\ref{adylam}), the change of variables becomes singular
for  $x_{12} \to 0$.
In this case, another rescaling of the variables
in (\ref{adylam}) gives
\eqabegin
  \O''_2 &=& \N\int d^2\la d^2c|\la|^{4j_1}|cx_{12}-\la|^{4j_2}
  \int d^2y\Phi_{j_1}(g(z_1),y)\Phi_{j_2}(g(z_2),y) \nn \\
   &=& -\N{\pi^4\o(2j_1+1)^2} i\cob(j_1+j_2+1)\cob^2(x_{12})
  \int d^2y\Phi_{j_1}(g(z_1),y)\Phi_{j_2}(g(z_2),y)
  \label{O2j-j-1} \label{O'2g}
  \comma
\eqaend
where we have used (\ref{GS}). Such a contribution should be added
to $\O'_2$ and we have $\O_2=\O'_2+\O''_2$.

Using the generators of $SL(2,\bfC)$ in (\ref{Da}), one can show that
the possible $SL(2,\bfC)$ invariant combinations of $x_a$ are only
$ i\delta(j_1-j_2) |x_{12}|^{4j_1} $ and
$ i\delta(j_1+j_2+1) \delta^2(x_{12})$
\cite{Teschner1}.
Thus we do not have any other invariants besides
(\ref{O2g}) and (\ref{O'2g}).
The appearance of the contact term (\ref{O'2g}) is
one of the special features of the $H_3^+$ WZW model: it is possible
since the left and right movers are combined from the beginning.

The projections in other cases are performed
similarly to the $n=2$ case.  For generic spins, we then obtain
\eqabegin
  \O_3(g(z_a)) & = &  \hf\N\prod_{a<b}^3|x_{ab}|^{2j_{ab}}
   \int\prod_{a=1}^3d^2y_a\Phi_{j_a}(g(z_a),y_a)
    \prod_{a<b}^3|y_{ab}|^{-2-2j_{ab}} \comma \label{O3O4} \\
  \O_4 (g(z_a)) &= &  \hf\N\prod_{a<b}^4\lf|x_{ab}\ri|^{-{2\o3}J+2(j_a+j_b)}
   \int_{y = x} {d^2y_1d^2y_2d^2y_3\o|y_{12}|^2|y_{23}|^2|y_{31}|^2}
    \prod_{a=1}^4\Phi_{j_a}(y_a)
  \prod_{a<b}^4\lf|y_{ab}\ri|^{{2\o3}J-2(j_a+j_b)}
  \period \nn
\eqaend
Here,
$J=\sum_{a=1}^4j_a$;  $x$ and $y$ are the cross-ratios
$x= x_{13}x_{24}/x_{14}x_{23}$ and
$y = y_{13}y_{24}/y_{14}y_{23}$ respectively; and
\eqabegin
  y_4={y_1y_2(1-x)+y_3(y_1x-y_2)\o y_1-y_2x+y_3(x-1)}
  \period
\eqaend
The factor $1/2$ on the right hand side of (\ref{O3O4})
needs some explanation.
For the change of the integration variables from $(a,c,d)$ in (\ref{acd})
to $y_a~(a=1,2,3)$, the Jacobian gives the factor $1/4$.
However, because this change of
variables is 2 to 1, we should multiply the integral by 2.

One can explicitly confirm that these are invariant under (\ref{RAPhi}).
For some special cases, possibly other invariants similar to
$\O''_2$ may appear.
Note that $\O_3$ and $\O_2$ are related by
\eqabegin
\lim_{j_3\riya -0}\O_3(g(z_a))=\O_2(g(z_a)) \period
\eqaend
Here, $j_3$ should approach zero from the negative real axis
so that the integral is convergent.

\mysubsection{Continuation in $j$}
Given the definition (\ref{Phicrfn}) and the invariants $\O_n$,
we would like to calculate the correlation functions of generic primary
fields $\Phi_j$. In this paper, we obtain them from the correlation
functions for $j \in \bfZ_{\geq 0}/2$ 
by continuing $j$ to generic values. 
(We also use some consistency conditions 
to determine the two-point function.)
The reason is two-fold. First, although
$\Phi_j$ is expanded by polynomials in $\gamma$ and
$e^{\pm \phi}$ for $j \in \bfZ_{\geq 0}/2$,
it becomes an infinite series for a generic $j$ when expanded
in $\gamma$ and $e^{\phi}$.
In such a case, it is not clear if the classical
expression (\ref{Phijcl}) makes sense in the quantum theory.
Second, it turns out that the explicit calculation is possible for
$j \in \bfZ_{\geq 0}/2$, since it is reduced to that in section 3.

This prescription may be justified by defining $\Phi_j$
as
\eqabegin
  \Phi_j &= & {1\o\Ga(-2j)}\int_0^{\infty} dt
     \,t^{-2j-1}\exp\big(-t\Phi_{\hf}\big) \nn \\
   &= &{1\o\Ga(-2j)}\int_0^{\infty} dt
   \,t^{-2j-1}\sum_{n=0}^{\infty}{(-1)^n\o n!}t^n\Phi_{n\o2}
  \comma \label{quantumPhij}
\eqaend
for generic $j \notin \bfZ_{\geq 0}/2$.
In this expression, we assume that all of the operators are regularized by the
point splitting method. The renormalized operator is discussed later.
From this definition, the invariant part
of $\prod_{a=1}^{n}\Phi_{j_a}(g(z_a),x_a)$ is given by
\eqabegin
    && \O_n(g(z_a),j_a,x_a) =  \int_0^\infty  \prod_{a=1}^n dt_a
   \, \frac{t^{-2j_a-1}}{\Ga(-2j_a)} \sum_{m_a=0}^{\infty}
    \lb \prod_a {(-1)^{m_a}\o m_a!} t_a^{m_a} \rb
   \O_n(g(z_a),\frac{m_a}{2},x_a)
  \period  \qquad \label{contOn}
\eqaend
Furthermore, for an analytic function $f(x)$ its value at a generic
point can be reconstructed from the data on the non-negative integers:
\eqabegin
  && {1\o\Ga(-x)}\int_0^{\infty} dt\,t^{-x-1}\sum_{n=0}^{\infty}{(-1)^n\o
n!}
  t^n f(n) \nn \\
  &=&{1\o\Ga(-x)}\int_0^{\infty} dt\,t^{-x-1}\sum_{n=0}^{\infty}{(-1)^n\o
n!}
  t^n \int_0^{\infty} ds e^{-ns}\til{f}(s) \label{contfn} \\
  &=&\int_0^{\infty} ds \til{f}(s)
 {1\o\Ga(-x)}\int_0^{\infty} dt\,t^{-x-1}e^{-te^{-s}} \nn \\
  &=& \int_0^{\infty} ds e^{-xs}\til{f}(s) \nn \\
  &=&  f(x)
  \period \nn
\eqaend
Here, $\til{f}(s)$ is the inverse Laplace transform of $f(x)$.
From (\ref{contOn}) and (\ref{contfn}),
we find that $ \Big\bra \O_n(g(z_a),j_a,x_a) \Big\ket $ is obtained by the
analytic continuation
\eqabegin
   \Big\bra \O_n(g(z_a),j_a,x_a) \Big\ket &=&
    \Big\bra \O_n(g(z_a),\frac{m_a}{2},x_a) \Big\ket \Bigm{|}_{m_a = 2j_a}
  \period   \label{contfinal}
\eqaend
This expression holds also when some of the spins $j_a$ are non-negative
half-integers. Thus the definition (\ref{quantumPhij})
indeed gives our prescription for the correlation functions.
We will shortly see that (\ref{quantumPhij}) is also consistent
with the renormalization.

If it is possible to calculate $\Big\bra \O_n(g(z_a),m_a/2,x_a) \Big\ket $
for arbitrary $m_a \in \bfZ_{\geq 0}$, we could take (\ref{quantumPhij})
as the starting point of the discussion for generic $j$. However,
we need to impose some conditions on $m_a$ in the later calculation.
Thus, our modest statement is that the continuation of the correlation
functions, the definition of the primary fields $\Phi_j$ in 
(\ref{quantumPhij}) 
and the renormalization are all consistent.

In any case, what we need in the following is to calculate
the correlation functions for the cases where the spins are non-negative
half-integers and then continue the results to other cases.
This is in the same spirit as that in \cite{GL} for the Liouville theory.
However, we remark that in our case there exists a parameter
region of $j$ in which the correlation functions are actually calculable.

\mysubsection{Renormalization}
Once we focus on the case $j \in \bfZ_{\geq 0}/2$,
the following calculation is carried out similarly to that in section 3.
As noticed there, the renormalization picks up a term which has
the strongest divergence. Such a term is obtained by dropping  
$e^{-\phi}$ in $\Phi_j$. 
For example, for $\O'_2$ in (\ref{O2g})
the surviving term is 
\eqabegin
  \til{\O}'_2(g(z_a)) & = &
   \pi^2\N|x_{12}|^{4j_2}i\cob(j_1-j_2)
       e^{2j_1(\phi(z_1)+\phi(z_2))} \nn \\
  && \quad \times
   \int d^2y_1d^2y_2|y_{12}|^{-4-4j_1}|y_1-\ga_1|^{4j_1}|y_2-\ga_2|^{4j_1}
   \period
\eqaend
Here and in the following, it is understood that integrals such as
the above are defined by the continuation from the parameter region
in which they converge. By simple changes of integration variables,
we further obtain
\eqabegin
  \til{\O}'_2(g(z_a)) & = & \pi^2\N a_2 i\cob(j_1-j_2)
     \Big(e^{\phi(z_1)+\phi(z_2)}|\ga_{12}|^2|x_{12}|^2\Big)^{2j_1}
      \nn \\
   &=&\pi^2\N a_2 i\cob(j_1-j_2) |x_{12}|^{4j_1} \tilde{\O}_2^G(j_a)
  \comma \label{tilO2}
\eqaend
where
\eqabegin
   a_2 & = & \int d^2y_1d^2y_2|y_{12}|^{-4-4j_1}|y_1|^{4j_1}|y_2-1|^{4j_1}
  \period
\eqaend
To compute $a_2$, we should regularize the integral to define it:
\eqabegin
a_2 &=&\lim_{\ep\riya 0}
\int d^2y_1d^2y_2|y_{12}|^{-4-4j_1+\ep}|y_1|^{4j_1}|y_2-1|^{4j_1} \nn \\
&=&\lim_{\ep\riya 0}\int d^2y_1|y_1-1|^{-4-4j_1+\ep}|y_1|^{4j_1}
\int d^2y_2|y_2|^{-2+\ep}|y_2-1|^{4j_1} \nn \\
&=& -{\pi^2\o (2j_1+1)^2}
\period
\label{a2}
\eqaend
Here, we have used the formula (\ref{singleint}). The first and second
integrals in the second line of (\ref{a2}) are zero and divergent,
respectively,  but the product gives a finite answer.
Note that $\til{\O}'_2$ is the same up to a factor as the invariant
combination
$\til{\O}_2^G$ appeared in section 3.
Thus the renormalization of $\O'_2$ is also given by
(\ref{O2ren}). It is straightforward to check that the terms in (\ref{O2g})
other than $\til{\O}'_2$ disappear after the renormalization.

If we simply apply the above argument to $\O''_2$ and  
drop the term $e^{-\phi}$ in $\Phi_j$, we obtain 
\eqabegin
 \til{\O}_2'' &=& -\N {\pi^4\o (2j_1+1)^2} i\cob(j_1+j_2+1)
 \cob^2(x_{12})e^{2j_1\phi(z_1)+2j_2\phi(z_2)}\int d^2y
 |y-\ga_1|^{4j_1}|y-\ga_2|^{-4-4j_1} \nn \\
  &=&  \N {\pi^6\o (2j_1+1)^4} i\cob(j_1+j_2+1)
 \cob^2(x_{12})\cob^2(\ga_{12})e^{2j_1\phi(z_1)+2j_2\phi(z_2)} \period
\label{tilO2'}
\eqaend
However, it is not clear if $\til{\O}''_2$ represents 
the correct contribution from $\O''_2$ after the renormalization: 
since $\O''_2$ always includes spins $j \notin \bfZ_{\geq 0}/2 $,
the above argument for $j \in \bfZ_{\geq 0}/2$ may not be valid. 
Here, we assume that $\O''_2$ is also renormalized by the multiplicative 
factor in (\ref{O2ren}). 
We do not use this expression in the actual calculation 
of $\Big\bra\til{\O}''_2\Big\ket$. Instead we determine it from 
consistency as in subsection 5.1.  

For the three-point function with $j \in \bfZ_{\geq 0}/2 $, 
the relevant term after the renormalization is obtained similarly to 
the case of $\O'_2$:
\eqabegin
   \til{\O}_3(g(z_a))&=&
   \hf\N\prod_{k<l}^3|x_{kl}|^{2j_{kl}}\int\prod_{a=1}^3d^2y_a
   e^{2j_a\phi_a}|y_a-\ga_a|^{4j_a}\prod_{a<b}^3|y_a-y_b|^{-2-2j_{ab}} \nn
\\
  &=& \hf\N a_3\prod_{a<b}^3\Big(e^{\phi_a+\phi_b}
   |\ga_{ab}|^2|x_{ab}|^2\Big)^{j_{ab}}  \label{tilO3} \\
  &=& \hf\N a_3 \prod_{a<b}^3 |x_{ab}|^{2j_{ab}} \til{\O}_3^G
  \period \nn
\eqaend
Here, the coefficient $a_3$ is 
\eqabegin
  a_3&=&\int\prod_{a=1}^3d^2y_a
  |y_1|^{4j_1}|y_2-1|^{4j_2}\prod_{a<b}^3|y_a-y_b|^{-2-2j_{ab}} \nn \\
  &=& \pi^3\lap(-N-1)\prod_{a=1}^3{\lap(2j_a-N)\o\lap(-2j_a)}
  \comma \label{a3}
\eqaend
with $N \equiv j_1 + j_2 + j_3$ and 
\eqabegin
  \lap(x) & = &{\Ga(x)\o \Ga(1-x)}
  \period
\eqaend
We notice again that $\til{\O}_3$ is $\til{\O}_3^G$ up to a factor,
and hence the renormalization is the same as that for $\til{\O}_3^G$.

These examples show that the divergence in the calculation is absorbed
by the renormalization of the primary fields,
\eqabegin
   \Phi_{\frac{n}{2}}^{\rm ren} &=& \epsilon^{-2\Delta_{n/2}}
  \Phi_{\frac{n}{2}}
  \period \label{Phiren}
\eqaend
This expression makes sense even for a generic spin $j$ through
(\ref{quantumPhij}). To see this, we first rewrite the renormalization
factor as
\eqabegin
   \epsilon^{-2\Delta_{j}} &=& \epsilon^{2b^2 j(j+1)} \ = \
        \alpha^j \int_{-\infty}^\infty \frac{d\lambda}{\sqrt{\pi}}
     \, e^{-\lambda^2 + 2 j \lambda \sqrt{\ln \alpha}}
    \comma
\eqaend
with $\alpha = \epsilon^{2b^2}$. Then, by rescaling the integration variable
$t$, we find that
\eqabegin
  \epsilon^{-2\Delta_j} \Phi_j
   &= &{1\o\Ga(-2j)}\int_0^{\infty} dt
   \,t^{-2j-1}\sum_{n=0}^{\infty}{(-1)^n\o n!}t^n
   \epsilon^{-2\Delta_{n/2}}\Phi_{n\o2}
   \period
\eqaend
Thus, (\ref{Phiren}) indeed absorbs the divergence for a generic $j$.
From now on, $\Phi_j$ should be understood to be the renormalized operator. 

Such a prescription for renormalization of operators may be reliable for the cases 
in which the expression eq.(\ref{quantumPhij}) can be used for the calculation, 
for example, in calculating $\Big\bra\O'_2\Big\ket$. Here we assume the same prescription of 
renormalization for other cases. We will confirm its validity by checking the 
relations among various correlation functions and also by finding an agreement with   
the results obtained in other approaches. 
\newpage
\mysection{Two- and three-point functions}
We are now ready to go into the details of the calculation. In this section,
we explicitly calculate the two- and three-point functions and
obtain them in closed forms. According to the argument in the previous
section, the spins are supposed to be non-negative half-integers
in the calculation of $ \Big\bra \til{\O}'_2 \Big\ket$ and 
$\Big\bra \til{\O}_3 \Big\ket$ until we arrive at the final expression. 
$\Big\bra \O''_2 \Big\ket$ is determined by some consistency conditions. 
We then carry out the 
analytic continuation in $j$ and obtain the results for generic spins.
\mysubsection{Two-point functions}
First, let us consider the two-point function
\eqabegin
   \Big\bra \Phi_{j_1}(g(z_1),x_1) \Phi_{j_2}(g(z_2),x_2) \Big\ket
    &=& \Big\bra \til{\O}_2' \Big\ket + \Big\bra \O''_2 \Big \ket
    \comma \label{twopt}
\eqaend
with $\Big\bra \til{\O}'_2 \Big\ket$ and $ \Big\bra \O''_2 \Big\ket$ 
given in (\ref{tilO2}) and (\ref{O'2g}), respectively.
Following the procedure in section 3, $\Big\bra \til{\O}_2^G \Big\ket$
in the first term is expressed by the integral \cite{Gawedzki}
\eqabegin
   \Big\bra \til{\O}_2^G(j) \Big\ket
    \ = \ && \Gamma(2j+1)
    |(z_0-z_1)(z_0-z_2)|^{4jb^2}|z_{12}|^{4j-4j^2b^2}  \\
   && \times \int \prod_{a=1}^{2j}
    \lb \frac{d^2 y_a}{\pi k}
   |y_a-z_0|^{-4b^2} |(y_a-z_1)(y_a-z_2)|^{-2+4jb^2} \rb
  \prod_{a<b}^{2j} |y_a-y_b|^{-4b^2}
  \period \nn
\eqaend
By making simple changes of variables\footnote{
One may take $z_0 = \infty$ here but this is not necessary.
}
and using the Dotsenko-Fateev formula (\ref{DFformula}) \cite{DF},
we obtain
\eqabegin
   \Big\bra \til{\O}_2^G(j) \Big\ket   &=&
      C_2(j)|z_{12}|^{4j(j+1)b^2}
      \comma
\eqaend
with
\eqabegin
C_2(j) &=& \Gamma(2j+1)
   \int \prod_{a=1}^{2j}
    \lb \frac{d^2 y_a}{\pi k} |y_a(y_a-1)|^{-2+4jb^2} \rb
   \prod_{a<b}^{2j} |y_a-y_b|^{-4b^2} \nn \\
&=&- kb^4 (2j+1)^2 \lf({\lap(b^2)\o k}\ri)^{2j+1}\lap(-(2j+1)b^2)
\period
\eqaend
$C_2(j)$ satisfies $C_2(0) =1$ and $\Big\bra \til{\O}_2^G(j) \Big\ket$
is independent of $z_0$ as it should be.
This has poles at $2j+1 = n(k-2)$  with positive
integers $n$. In particular, the first one from $n=1$ corresponds
to the convergence
condition discussed in \cite{Gawedzki}, which was associated with the fusion
rule $j \leq \til{k}/2 \equiv (k-4)/2$ of the $SU(2)$ WZW model.
Combining $C_2(j)$ with $a_2$ yields
\eqabegin
   B(j) &\equiv & \pi^2\N a_2 C_2(j)
    \ =  \ \N\pi^4kb^4\lf({\lap(b^2)\o k}\ri)^{2j+1}\lap(-(2j+1)b^2)
   \period
\eqaend

Next, we turn to the contribution from $ \Big\bra \O''_2 \Big\ket$. 
Since this term
always has $j \notin \bfZ_{\geq 0}/2 $ and  
may contain distributions such as $\delta^2(\gamma_{12})$ in (\ref{tilO2'}), 
we do not know how to calculate its
expectation value in our formalism. However, it is determined by
the consistency as follows.

First, let us recall that $\Phi_j$ and $\Phi_{-j-1}$ are classically
related by (\ref{j-j-1}). The structure of the integral transformation
is almost completely fixed by the $SL(2,\bfC)$ symmetry. Here, we allow
that the coefficient in front of the integral changes in the quantum
theory as
\eqabegin
  \Phi_j(g,x) & = & R(j)\int d^2y|x-y|^{4j}\Phi_{-j-1}(g,y)
  \period \label{reflection}
\eqaend
$R(j)$ is the reflection coefficient.
Repeating the above transformation twice gives
\eqabegin
  R(j)R(-j-1) & = & - {(2j+1)^2 \o \pi^2}
   \period \label{RjR-j-1}
\eqaend
Next, by introducing $A(j)$ to denote the coefficient in
$\Big\bra \O''_2 \Big\ket$,
we rewrite (\ref{twopt}) as
\eqabegin
  && \Big\bra\Phi_{j_1}(g(z_1),x_1)\Phi_{j_2}(g(z_2),x_2)\Big\ket
    \label{twoptfinal} \\
  &=&|z_{12}|^{4b^2j_1(j_1+1)}\Big[A(j_1) i\cob(j_1+j_2+1)\cob^2(x_{12})
   +B(j_1) i\cob(j_1-j_2)|x_{12}|^{4j_1}\Big]
   \period \nn
\eqaend
Substituting (\ref{reflection}) into $\Phi_{j_1}$, we obtain
\eqabegin
   A(j) &=& - \frac{\pi^2}{(2j+1)^2} R(j) B(-j-1) \comma  \nn \\
   B(j) &=& R(j) A(-j-1)
    \period \label{ABR1}
\eqaend
Further substitution of (\ref{reflection}) into $\Phi_{j_2}$
gives
\eqabegin
   A(j) &=& A(-j-1) \comma \nn \\
   B(j) &=& - \frac{\pi^2}{(2j+1)^2} R^2(j) B(-j-1)
    \period \label{ABR2}
\eqaend
Together with the result of $B(j)$, these determine $A(j)$ and $R(j)$:
\eqabegin
  A(j)&=&   -\N {\pi^5kb^2\o (2j+1)^2} \comma \nn \\
  R(j)&=& -{(2j+1)^2b^2\o \pi}\lf({\lap(b^2)\o k}\ri)^{2j+1}
        \lap(-(2j+1)b^2)
  \period
\eqaend
Since the consistency conditions (\ref{ABR1}) and (\ref{ABR2}) are
invariant under $(A(j),B(j),R(j))\riya (-A(j),B(j),-R(j))$,
there is an ambiguity in the
sign of $A(j)$ and $R(j)$ for a given $B(j)$.
This sign is fixed by demanding
that $R(j)$ is reduced to its classical value $(2j+1)/\pi$
in the limit $k\riya\infty$.
This completes the calculation of the two-point function.
\newpage
\mysubsection{Three-point functions}
Let us move on to the discussion of the three-point function.
In the previous section, we argued that for generic $j$ the three-point
function is given by
\eqabegin
 &&  \Big\bra \Phi_{j_1}(g(z_1),x_1) \Phi_{j_2}(g(z_2),x_2)
       \Phi_{j_3}(g(z_3),x_3) \Big\ket  \nn \\
 && \qquad \ = \ \Big\bra \til{\O}_3 (g(z_a),x_a) \Big\ket
     \ \equiv \ \hf\N a_3 \prod_{a<b}^3 |x_{ab}|^{2j_{ab}} G_3(j_a,z_a)
  \comma
\eqaend
with $G_3 \equiv \Big\bra \til{\O}^G_3 \Big\ket$.
Following the procedure in section 3,
for the spins satisfying (\ref{j1-j3}),
$G_3$ is given by the integral
\eqabegin
   G_3 &=&
   \prod_{a<b}^3 |z_a-z_b|^{-4j_aj_b b^2 + 2 j_{ab}}
   \prod_{a=1}^3 |z_0-z_a|^{4j_ab^2}
   \int \prod_{a=1}^N \frac{d^2y_a}{\pi k} \ |y_a-z_0|^{-4b^2}
   \nn \\
   && \times
       \prod_{b=1}^3 |y_a-z_b|^{4j_a b^2}
     \prod_{a<b}^N |y_a-y_b|^{-4b^2}
     \sum_{\sigma \in S_N}
   \prod_{a \leq 2j_1} \lbb(y_a-z_1)(\bar{y}_{\sigma(a)}-\bz_1)\rbb^{-1}
   \\
  && \times
   \prod_{a \leq j_{12} \ {\rm or} \ a>2j_1}
   \lbb(y_a-z_2)(\bar{y}_{\sigma(a)}-\bz_2)\rbb^{-1}
   \prod_{a > j_{12}} \lbb(y_a-z_3)(\bar{y}_{\sigma(a)}-\bz_3)\rbb^{-1}
  \comma \nn
\eqaend
where $ S_N$ stands for the permutations
of $N=j_1+j_2+j_3$ elements. Note that $2j_2 \geq j_{12}$.
By some changes of variables, this is brought into the form
\eqabegin
  G_3 & = & \prod_{a<b}^3|z_{ab}|^{2\Delta_{ab}} C(j_1,j_2,j_3,\xi)
  \comma
\eqaend
where
$\xi$ is the cross-ratio
\eqabegin
  \xi={z_{01}z_{23}\o z_{03}z_{21}}
  \comma
\eqaend
and $\Delta_{ab}$ are given by
\eqabegin
  \Delta_{12} & = & \lap_{j_3}-\lap_{j_1}-\lap_{j_2} \ = \
   b^2[j_{12}(N+1)-2j_1j_2]
   \comma
\eqaend
and similar expressions.
The coefficient $C(j_1,j_2,j_3,\xi)$ is roughly speaking a kind of a four-point function,
\eqabegin
\int e^{-S_{WZW}}\Phi_{j_1}(0)\Phi_{j_2}(1)\Phi_{j_3}(\infty)   \h{\Phi}_0(\xi),
\eqaend
 and we can obtain
\eqabegin
  & & C(j_1,j_2,j_3,\xi)  \\
 && =
 |\xi|^{4b^2j_1}|1-\xi|^{4b^2j_2}\int \prod_{a=1}^N{d^2y_a\o \pi k}
 |y_a|^{4b^2j_1}|y_a-1|^{4b^2j_2}|y_a-\xi|^{-4b^2}
 \prod_{a<b}|y_a-y_b|^{-4b^2} \nn \\
&& \quad \times \sum_{\si\in S_N}\prod_{a\leq 2j_1}{1\o y_a\b{y}_{\si(a)}}
   \prod_{a\leq j_{12}}{1\o (y_a-1)(\b{y}_{\si(a)}-1)}\prod_{a>2j_1}
   {1\o (y_a-1)(\b{y}_{\si(a)}-1)}
  \period \nn
\eqaend
Since $j_a$ $(a=1,2,3)$ were on an equal footing originally,
different changes of variables in $G_3$ give 
expressions in which $j_a$ are permuted:
\eqabegin
  && C(j_1,j_2,j_3,\xi)=C(j_2,j_1,j_3,1-\xi)=C(j_3,j_2,j_1,{1\o \xi})
  \period
\label{Cxi}
\eqaend

Since $\Phi_j$ are conformal fields, their correlation function
factorizes into the holomorphic and anti-holomorphic parts.
Therefore, if $C(j_a,\xi)$ has no singularity in $\xi$,
it is an entire function on $\CP^1$, a constant.
The possible singularities of $C(j_a,\xi)$ are located at the insertion
points of other operators, i.e.,
$\xi=0,1,\infty$.
From the relation (\ref{Cxi}), the existence of the limit
\eqabegin
  \lim_{\xi\riya 0}C(j_1,j_2,j_3,\xi) & \equiv & C_3(j_1,j_2,j_3)
  \comma \label{Clim}
\eqaend
ensures that $C(j_a,\xi)$ is independent of $\xi$,  as it should be.
We show this
by an explicit calculation.

For this purpose, we make the rescalings of variables
\eqabegin
  y_a & = & \xi w_a, \qquad(a=1,\ldots,2j_1)
  \period
\eqaend
We can then take the limit $\xi \to 0 $ and find that many terms drop out.
Consequently, we obtain
\eqabegin
  C_3(j_a) &=& \Ga(2j_1+1)\Ga(j_{23}+1)
       \int \prod_{a\leq 2j_1} \frac{d^2w_a}{\pi k}
  \, |w_a|^{4b^2j_1-2} |w_a-1|^{-4b^2}
     \prod_{a<b \leq 2j_1} |w_a-w_b|^{-4b^2} \nn \\
   && \times
   \int \prod_{a > 2j_1} \frac{d^2y_a}{\pi k}
   \, |y_a|^{-4(j_1+1)b^2} |y_a-1|^{4j_2b^2-2}
     \prod_{2j_1<a<b} |y_a-y_b|^{-4b^2}
    \period \label{C3j1}
\eqaend
Here, we have used the following equations:
\eqabegin
& &\lim_{\xi\riya 0}\prod_{a=1}^Nd^2y_a
\sum_{\si\in S_N}\prod_{a\leq 2j_1}{1\o y_a\b{y}_{\si(a)}}
\prod_{a\leq j_{12}}{1\o (y_a-1)(\b{y}_{\si(a)}-1)}\prod_{a>2j_1}
{1\o (y_a-1)(\b{y}_{\si(a)}-1)} \nn \\
&=&\prod_{a\leq 2j_1}d^2w_a\prod_{a>2j_1}d^2y_a
\sum_{\si\in S_{2j_1}}\sum_{\tau\in S_{j_{23}}}\prod_{a\leq 2j_1}
{1\o w_a\b{w}_{\si(a)}}\prod_{a>2j_1}{1\o (y_a-1)(\b{y}_{\tau(a)}-1)}
  \\
 &=&\Ga(2j_1+1)\Ga(j_{23}+1)\prod_{a\leq 2j_1}d^2w_a|w_a|^{-2}
  \prod_{a>2j_1}d^2y_a|y_a-1|^{-2}
  \period \nn
\eqaend

From (\ref{C3j1}), we find that $C_3(j_a)$ is factorized into
a product of two Dotsenko-Fateev integrals in (\ref{DFformula}):
\eqabegin
 C_3(j_a) & = & {1\o (\pi k)^N}\Ga(2j_1+1)\Ga(j_{23}+1)  \\
   && \qquad \times
  J_{2j_1}(2b^2j_1-1,-2b^2,b^2)J_{j_{23}}(-2b^2j_1-2b^2,2b^2j_2-1,b^2)
  \period \nn
\eqaend
The first integral is evaluated to be
\eqabegin
  J_{2j_1}(2b^2j_1-1,-2b^2,b^2) & = &
{\lf(\pi\lap(b^2)\ri)^{2j_1+1}\o\pi\Ga(2j_1+1)\lap((2j_1+1)b^2)}
   \period
\eqaend
The second integral takes the form $ J_{j_{23}} \sim
\prod_{n=1}^{j_{23}} \lbb \Delta(f_1(n)) \Delta(f_2(n))....\rbb$
with certain functions $f_a(n)$.
Although, as discussed in section 4, we would like to analytically continue
the final expression in term of $j_a$, it appears difficult to do so in this
form.
However, this is achieved with the help of the
$\Upsilon$-function introduced in \cite{ZZ} (see also \cite{DO}).
We have collected the definition and
basic properties of $\Upsilon(x)$ in appendix B.
To use $\Upsilon(x)$, we first rewrite
$J_{j_{23}}$ using $I_m(\alpha_a)$ defined in (\ref{Im}):
\eqabegin
  J_{j_{23}}(-2b^2j_1-2b^2,2b^2j_2-1,b^2)
 & = &I_{j_{23}}(j_1b+b,-j_2b+{1\o 2b},-j_3b+{1\o 2b})
  \period
\eqaend
By further making use of the relation between $\Upsilon(x)$ and $I_m $,
(\ref{ImUpsilon}), we arrive at
\eqabegin
  C_3(j_a) & = &
  kb\lf({1\o k}b^{-2b^2}\lap(b^2)\ri)^{N+1}{\up'(0)\o\up((N+2)b)}
 \prod_{a=1}^3{\up((2j_a+1)b)\o\up((N-2j_a+1)b)}
  \comma
\eqaend
where $\Upsilon'(x) = d \Upsilon/dx$.
This is symmetric with respect to $j_a$, though
the expressions in the intermediate stage were not.
Since $\Upsilon(x)$ is analytic in $x$, we can continue the above expression
to that for arbitrary $j_a$.
Finally, putting everything together, we obtain the expression of the
three-point function,
\eqabegin
   \lf\bra \prod_{a=1}^3 \Phi_{j_a}(g(z_a),x_a) \ri\ket
 &=&  D(j_a)\prod_{a<b}^3 |x_{ab}|^{2j_{ab}} |z_{ab}|^{2\lap_{ab}}
   \comma \label{threeptfinal}
\eqaend
with
\eqabegin
  D(j_a)& \equiv & \hf\N a_3C_3(j_a) \\
  &= & \hf\N \pi^3kb^4\lf({1\o k}b^{-2b^2}\lap(b^2)\ri)^{N+1}
  {\up'(0)\o\up(-(N+1)b)}
  \prod_{a=1}^3{\up(-2j_ab)\o\up((2j_a-N)b)}
 \period \nn
\eqaend

The structure of the poles of the three-point function
is important to consider the fusion
rules \cite{Teschner1,Teschner2}. It is read off
from the zeros of $\up(x)$ given in (\ref{zeros}).
For generic $j_a$, $D(j_a)$ has poles at
\eqabegin
  j_{ab} \comma N+1 &=& m  + n b^{-2} \comma \ -(m+1) - (n+1) b^{-2}\comma
   \quad m,n \in \bfZ_{\geq 0}
  \period
\eqaend
However, for example, for $j_a \in \bfZ_{\geq 0}/2$, many poles are
canceled with the zeros from the numerator. This is also confirmed 
by noting that
in this case $\up (x)$ is reduced to the Dotsenko-Fateev integral
given by a product of $\Delta(x)$.
Incidentally, the pole at $N=k-3 =\til{k}+1$ corresponds to
the convergence condition
discussed in \cite{Gawedzki}, which was associated to the three-point
fusion rule of the $SU(2)$ WZW model, $j_1+j_2+j_3 \leq \til{k}$.

Finally, let us check the consistency of our calculation.
As the simplest check, we see that $C_3(j,j,0) = C_2(j)$.
Furthermore, for $j_{1,2} = -1/2 + i \rho_{1,2}$ and $j_3 \to -0$,
the three-point function is reduced exactly to the two-point function:
\eqabegin
&&\lim_{j_3\riya -0}D(j_a)\prod_{a<b}|x_{ab}|^{2j_{ab}}
|z_{ab}|^{2\lap_{ab}} \nn \\
&=&
|z_{12}|^{-4\lap_{j_1}}\Big[A(j_1)i\cob(j_1+j_2+1)\cob^2(x_{12})
+B(j_1)i\cob(j_1-j_2)|x_{12}|^{4j_1}\Big]
 \period \label{3to2}
\eqaend
To derive this relation, we need to take the limit carefully
so that we do not miss the distributions \cite{Teschner2}.
Here, the factor $a_3$, which comes from the $SL(2,\bfC)$
projection, gives the delta functions of $j$'s.

\mysection{Comparison with other approaches}
In the previous section,
we obtained the closed forms of the two- and three-point functions
(\ref{twoptfinal}) and (\ref{threeptfinal}), which are valid for generic
spins. Now let us compare these with the results obtained by other
approaches.
%
\mysubsection{Supergravity approximation}
In the discussion of the AdS/CFT correspondence, the various
correlation functions have been calculated in the supergravity
approximation, e.g., \cite{GKP,Witten},\cite{MV}-\cite{KO}.
The supergravity approximation for the correlation functions of
$\Phi_j$ is nothing but the zero-mode approximation
from the point of view of the $H_3^+$ WZW model \cite{Teschner1}.

Note that $\Phi_j$ coincides with the boundary-to-bulk propagator
on $AdS_3$, i.e., $\Phi_j$ satisfies
\eqabegin
&&(\lap-m^2)\Phi_j(g,x)=0 \comma \nn \\
&&\lim_{\phi\riya\infty}\Phi_j(g,x)=-{\pi \o
2j+1}e^{-2(j+1)\phi}\cob^2(\ga-x)
\comma
\eqaend
where $\lap=-4\eta_{ab}J^a_0J^b_0/k$ is the Laplacian on $AdS_3$ and
the mass $m$ and spin $j$ are related by
\eqabegin
j=-\hf-\hf\rt{1+km^2} \period
\eqaend

Since the two-point function in the supergravity calculation
is not the same object as ours, we will not make a direct comparison.
In the case of the three-point function,  the supergravity result
\cite{Teschner1,FMMR} is obtained by following the so-called
GKPW-prescription \cite{GKP,Witten}:
\eqabegin
  \int e^{2\phi}d\phi
  d^2\ga\prod_{a=1}^3\Phi_{j_a}(g,x_a)
  & = & D(j_a)^{SG}\prod_{a<b}|x_{ab}|^{2j_{ab}}
  \comma \label{sugra}
\eqaend
with
\eqabegin
  D(j_a)^{SG} & = &{\pi\o2}\Ga(-N-1)\prod_{a=1}^3{\Ga(2j_a-N)\o\Ga(-2j_a)}
  \period \label{Dsg}
\eqaend
In our world-sheet calculation, the supergravity limit corresponds to
$\al'\riya 0$ or $k\riya \infty$.
Since $\up(x)$ appears singular in this limit, i.e., $b \to 0$
(see (\ref{Upsilon})), it is useful to go back to the expression using
$J_m$.  We then find that in this limit our calculation
is reduced to the supergravity approximation as expected:
\eqabegin
\lim_{k\riya\infty}D(j_a)& = & \N\pi^2D(j_a)^{SG}
\period \label{tosugra}
\eqaend

From (\ref{Dsg}),
we see that the basic structure of $D(j_a)^{SG}$, such as the location
of the poles, is encoded in the coefficient $a_3$ in (\ref{a3}).
This is because the zero-mode integral in (\ref{sugra}) is essentially
the same as the integral in the $SL(2,\bfC)$ projection
in (\ref{SL2projection}).
However, the precise connection is not yet clear.
The coefficient in (\ref{tosugra}) may indicate a
difference of these two integrals.

\mysubsection{Bootstrap approach}
Next, we turn to the other approach to the quantum theory.
In \cite{Teschner1,Teschner2},
the correlation functions for generic spins
are discussed based on the symmetry and bootstrap.
In particular, the three-point function is obtained as the solution
to the functional relation derived from the crossing symmetry.

Since the normalizations in \cite{Teschner1} and \cite{Teschner2}
are different, we first consider the result in \cite{Teschner2}.
The three-point function in \cite{Teschner2} corresponding to our $D(j_a)$
is\footnote{
We put an extra minus sign in the original expression in
\cite{Teschner2}. This sign is needed because
$j_3$ is taken to zero from the positive real
axis in \cite{Teschner2} to obtain the two-pint function from the
three-point function.
}
\eqabegin
 D(j_a)^T &=& -{1\o2\pi^3 b}\lf({\pi b^{2b^2-2}\o\lap(b^2)}\ri)^{N+2}
 {\up'(0)\o\up(-(N+1)b)}\prod_{a=1}^3{\up(-(2j_a+1)b)\o\up((2j_a-N)b)}
  \period
\eqaend
The two-point function is obtained
by taking the limit $j_3 \to -0$ with $j_a = -1/2 + i \rho_a$ $(a=1,2)$
as in (\ref{3to2}):
\eqabegin
  \lim_{j_3\riya -0}D(j_a)^T\prod_{a<b}^3|x_{ab}|^{2j_{ab}}
 & = &
  i\cob(j_1-j_2)B(j_1)^T|x_{12}|^{4j_1}+ i\cob(j_1+j_2+1)\cob^2(x_{12})
 \comma
\eqaend
where
\eqabegin
  B(j)^T & = &
   {1\o\pi b^2}\lf({\pi\o b^2\lap(b^2)}\ri)^{2j+1}\lap(-(2j+1)b^2)^{-1}
  \period
\eqaend
Once the above expression is obtained, one can continue it to generic
$j$. In this expression,
the quantity corresponding to our $A(j)$ is $A(j)^T=1$,
and the reflection coefficient in \cite{Teschner2} is given by
$ R(j)^T=B(j)^T$. We can check that these $A(j)^T$, $B(j)^T$ and $R(j)^T$
satisfy the consistency conditions (\ref{RjR-j-1}), (\ref{ABR1})
and (\ref{ABR2}).

By comparing these with ours, we find that
the results in \cite{Teschner2} are equivalent to ours,
since the difference is absorbed by the normalization of the primary fields.
To see this, we first note that, from the
two-point functions $B(j)$ and $B(j)^T$, our $\Phi_j$
and the primary fields $\Phi^T_j$ in \cite{Teschner2} are related by
\eqabegin
   \Phi_j= E(j)\Phi_j^T
   \comma \label{PhiPhiT}
\eqaend
with
\eqabegin
E(j) = \lf({B(j)\o B(j)^T}\ri)^{\hf}=-\N^{\hf}\pi^2b^4\lf({b^2\o \pi
k}\ri)^j
\lap(b^2)^{2j+1}\lap(-(2j+1)b^2) \period
\eqaend
This rescaling is consistent with the
relation between $A(j)$ and $A(j)^T$:
\eqabegin
 A(j)&=& A(j)^T E(j)E(-j-1)
  \period
\eqaend
Furthermore, the three-point functions satisfy
\eqabegin
  D(j_a)=\lf( {1\o \pi^4\N}\ri)^{\hf}D(j_a)^T\prod_{a=1}^3E(j_a)
  \period
\eqaend
Thus, with the choice of the normalization factor $\N$,
\eqabegin
 \N & = & {1\o\pi^4}
  \comma
\eqaend
the two results are in complete agreement
including the numerical coefficients.

In addition, the relation between the normalizations in \cite{Teschner1} and
\cite{Teschner2} has been discussed in \cite{Teschner2}.
From (\ref{PhiPhiT}), we find that our normalization is essentially
the same as that in \cite{Teschner1}.\footnote{
The normalization in\cite{Teschner1} is not completely fixed.
}
For generic $j_a$, the choice of these two normalizations is irrelevant
to the pole structure. However, it is relevant in some cases.

\mysection{Discussion}
Using the path integral approach, we discussed the correlation functions
of the primary fields in the $SL(2,\bfC)/SU(2)$ WZW model which
corresponds to the string theory on the Euclidean $AdS_3$.
Because of the non-compactness of $SL(2,\bfC)/SU(2) = H_3^+$,
a careful definition of the correlation functions was necessary.
We argued that the calculation for generic primary fields is reduced to
that of Gaw{\c e}dzki for certain invariants with non-negative
half-integral spins. The point was the $SL(2,\bfC)$ projection for
$\Phi_j$ in section 4.2 and the analytic continuation in the spin $j$.
Regarding the latter, there still remain subtleties and hence
we may need further discussions for a rigorous treatment.
We then carried out an explicit calculation of the two- and
three-point functions and obtained their closed forms.
The three-point function was reduced to the supergravity result
in the semi-classical limit. Furthermore, by an appropriate
change of the normalization of the primary fields, we found an exact
agreement with the results by Teschner using the bootstrap approach.
Notice that a mere analytic continuation of Gaw{\c e}dzki's correlation
functions $C_2,C_3$
does not reproduce the Teschner's results. The coefficients $a_2,a_3$
which were derived from the $SL(2,\C )$ projection were important.

As discussed in the introduction, the $H_3^+$ WZW model
has applications in various directions. The exact result of
the correlation functions will be used for these investigations.
Some applications are found in \cite{GK,MO}. In particular,
it will serve as the starting point for the precise understanding
of the AdS/CFT correspondence beyond the supergravity approximation.
It is also interesting to apply our formalism to
the correlation functions of other fields, such as the energy-momentum
tensor of the boundary CFT.

As for the $H_3^+$ WZW model itself, it is important to study the
fusion rules and the issue of the factorization in order
to know the true spectrum of the model.
From (\ref{L2}), the Hilbert space of the model consists of the principal
continuous series. Thus, the states in this series may give the complete
basis when one factorizes the four-point functions. On the other hand,
since the spins take continuous values, the poles in the three-point
function may contribute to the fusion rules and the states in
other representations may appear.
The role of such states seems to be similar to that of the non-normalizable
states in the Liouville theory.
These issues have also been discussed in \cite{Teschner1,Teschner2}.
In the case of the $SL(2,\bfR)$ WZW model corresponding to the Lorentzian $AdS_3$, 
it has been argued that the winding modes play an important role
\cite{MO}\cite{Maldacena:2000kv}\cite{Giribet:2000fy} (see also 
\cite{Henningson:1991jc}\cite{Satoh:1998xe}). It is not clear 
how to incorporate such modes in our formalism. 

In our formalism or that in \cite{Teschner1,Teschner2},
it seems difficult to push the calculation to the higher point
functions though it is possible in principle. The free field approach
discussed in \cite{CTT,GN} (and the free field approach to
$\widehat{sl_2}$) is certainly a powerful tool for this purpose.
As discussed in section 3, the path integral approach gives 
expressions which look very similar to those in the free field approach.
This implies that, when appropriately treated,
the free field approach might be used in the region
besides near the boundary of $H_3^+$. Thus, it will be useful to consider the
precise connection between these two approaches.

\vskip 10ex
\begin{center}
 {\sc Acknowledgments}
\end{center}
\par \smallskip

We would like to thank J. de Boer, G. Giribet, A. Giveon, K. Hamada,
K. Hori, K. Hosomichi, H. Ishikawa, K. Ito, K. Itoh, M. Kato, K. Mohri,
C. N{\'u}{\~n}ez, Y. Sugawara, N. Taniguchi for useful discussions
and correspondences.
The work of N.I. was supported in part by Grant-in-Aid for Scientific
Research from the Ministry of Education, Science and Culture in Japan.
The work of K.O. was supported in part by JSPS Research Fellowships for
Young Scientists. The work of Y.S. was also supported in part
by Grant-in-Aid for Scientific Research on Priority Area 707 from
the Ministry of Education, Science and Culture in Japan.
\newpage
\setcounter{section}{0}
%
\appsection{Integral formulas}
In this appendix we collect useful integral formulas.
The first one is the Dotsenko-Fateev formula \cite{DF}, given by
\eqabegin
  J_n(\al,\bt,\rho) & =
      &\int\prod_{i=1}^nd^2y_i|y_i|^{2\al}|y_i-1|^{2\bt}
  \prod_{i<j}|y_i-y_j|^{-4\rho} \nn \\
  &=& n!\pi^n\lf({\Ga(1+\rho)\o\Ga(-\rho)}\ri)^n\prod_{l=1}^n
   {\Ga(-l\rho)\o\Ga(1+l\rho)} \label{DFformula} \\
  && \times \prod_{l=0}^{n-1}{\Ga(1+\al-l\rho)\Ga(1+\bt-l\rho)
  \Ga(-1-\al-\bt+(n-1+l)\rho)
   \o \Ga(-\al+l\rho)\Ga(-\bt+l\rho)\Ga(2+\al+\bt-(n-1+l)\rho)}
  \period \nn
\eqaend
Setting $n=1$ and $\rho = 0 $ in the above, we obtain the second one,
\eqabegin
  \int d^2z|z|^{2\alpha}|z-1|^{2\beta} & = &
\pi\Delta\lf(1+\alpha\ri)\Delta\lf(1+\beta\ri)\Delta\lf(-1-\alpha-\beta\ri)
     \comma \label{singleint}
\eqaend
where $\Delta(x)= \Gamma(x)/\Gamma(1-x)$.
\appsection{$\Upsilon$-function}
Here, we give the definition of the $\up$-function and its basic properties.
The function $\up(x)$ is defined by \cite{ZZ} (see also \cite{DO})
\eqabegin
  \log\up(x ) & = & \int_0^{\infty}{dt\o t}\lf[({Q\o2}-x)^2e^{-t}-
 {\sinh^2\lf({Q\o2}-x\ri){t\o2}\o \sinh{bt\o2}\sinh{t\o 2b}}\ri]
  \comma \label{Upsilon}
\eqaend
with $Q=b+1/b$. This integral converges in the strip $ 0 < $ Re $ x < Q$.
For other $x$, $\up(x)$ is defined through
the functional relations
\eqabegin
  &&\up(x+b)=\lap(bx)b^{1-2bx}\up(x) \comma \nn \\
  &&\up\Big(x+{1\o b}\Big)=\lap\lf({x\o b}\ri)b^{{2x\o b}-1}\up(x)
   \comma \label{+b}\\
  &&\up(Q-x)=\up(x)
  \period \nn
\eqaend
From these relations,
one finds that $\up(x)$ has zeros at
\eqabegin
   && x=-mb-{n\o b},\quad (m+1)b+{n+1\o b}, \qquad m,n\in \Z_{\geq 0}
   \period \label{zeros}
\eqaend

This function can be used in the analytic continuation
of the
the Dotsenko-Fateev integral $J_n$ in (\ref{DFformula}) with respect
to $n$. To see this, we first denote a Dotsenko-Fateev integral by
$I_m(\alpha_a)$:
\eqabegin
 I_m(\al_1,\al_2,\al_3) & = & J_m(-2b\alpha_1,-2b\alpha_2,b^2)
  \label{Im} \\
   &=&
   \Gamma(m+1) \lb \pi \Delta(1+b^2)\rb^m \prod_{l=1}^m \Delta(-lb^2)
   \prod_{l=0}^{m-1}\prod_{a=1}^3 \Delta^{-1}(2b\alpha_a+lb^2)
  \comma \nn
\eqaend
where $\al_3$ is given by
\eqabegin
 \Sigma & \equiv &  \sum_{i=1}^3\al_i \ = \ Q-mb
 \period
\eqaend
Using (\ref{+b}), this can be rewritten as \cite{ZZ,ORPS}
\eqabegin
 I_m(\al_a)&= & \Ga(m+1)\lf(-\pi
b^{2-2b^2}\lap(b^2)\ri)^m{\up'(0)\o\up'(-mb)}
\prod_{i=1}^3{\up(2\al_i)\o\up(\Si-2\al_i)} \nn \\
&=&\lf(\pi b^{-2b^2}\lap(b^2)\ri)^m
{\up'(0)\o\Ga(m+1)\up((m+1)b)}\prod_{i=1}^3{\up(2\al_i)\o\up(\Si-2\al_i)}
 \period \label{ImUpsilon}
\eqaend
Here, $\up'(x) = d\up/dx$ and we have used the relation
\eqabegin
 \up'(-mb)&=&(-1)^mb^{2m}\up((m+1)b)\Ga(m+1)^2
  \period
\eqaend
In particular, we have
\eqabegin
  \up'(0) &=& \up(b)
  \period
\eqaend
%
%
%
\def\thebibliography#1{\list
 {[\arabic{enumi}]}{\settowidth\labelwidth{[#1]}\leftmargin\labelwidth
  \advance\leftmargin\labelsep
  \usecounter{enumi}}
  \def\newblock{\hskip .11em plus .33em minus .07em}
  \sloppy\clubpenalty4000\widowpenalty4000
  \sfcode`\.=1000\relax}
 \let\endthebibliography=\endlist
%
%
\vskip 10ex
\begin{center}
 {\sc References}
\end{center}
\par
%

%

\end{document}